\documentclass[aps,pre,twocolumn,amsmath,superscriptaddress,floatfix]{revtex4-1}
\usepackage[latin1]{inputenc}
\usepackage[T1]{fontenc}
\usepackage{amsmath}
\usepackage{amsfonts}
\usepackage{amssymb}
\usepackage{bm}
\usepackage{mathtools}
\usepackage{graphicx}
\usepackage{bigints}

\usepackage[dvipsnames]{xcolor}
\usepackage[normalem]{ulem}

\begin{document}

\title{Hamiltonian derivation of the point vortex model from the two-dimensional nonlinear Schr{\"o}dinger equation}

\author{Jonathan Skipp}
\affiliation{Department of Mathematics, College of Engineering and Physical Sciences, Aston University, Aston Triangle, Birmingham, B4 7ET, UK}

\author{Jason Laurie}
\affiliation{Department of Mathematics, College of Engineering and Physical Sciences, Aston University, Aston Triangle, Birmingham, B4 7ET, UK}

\author{Sergey Nazarenko}
\affiliation{Universit\'{e}  C\^{o}te d'Azur,   CNRS, Institut de Physique de Nice, Parc Valrose, 06108 Nice, France}

\begin{abstract}
We present a rigorous derivation of the point vortex model starting from the two-dimensional nonlinear Schr{\"o}dinger equation, from the Hamiltonian perspective, in the limit of well-separated, subsonic vortices on the background of a spatially-infinite strong condensate. As a corollary, we calculate to high accuracy the self-energy of an isolated elementary Pitaevskii vortex, for the first time. 
\end{abstract}

\maketitle

\section{Introduction}

In two-dimensional (2D) cold-atom systems such as superfluids and Bose gases, as well as in nonlinear optical systems, the dynamics can often be modelled by a field $\psi(\mathbf{x},t) : \mathbb{R}^2\times (0,\infty) \mapsto \mathbb{C}$ (physical examples of which we specify in Sec.\ \ref{subsec:NLSphysical} below) evolving via the 2D nonlinear Schr\"odinger (NLS) equation \citep{sulem2007nonlinear},
\begin{equation}
	\label{eq:NLS}
	i\frac{\partial\psi}{\partial t} + \nabla^2 \psi - |\psi|^2\psi + \psi = 0,
\end{equation} 
where $\nabla^2 = \partial_x^2 + \partial_y^2$.
In an infinite domain, the stationary ground state solution of Eq.\ \eqref{eq:NLS} has constant density $\rho=|\psi|^2$ everywhere (for convenience we have normalised the density to $\rho_0=1$, see Sec.\ \ref{subsec:NLSphysical}).
	This state is known as the uniform condensate solution.

	An important dynamical regime manifested by the 2D NLS equation is that of a strong condensate punctuated by $N$ well-separated, subsonic (so that compressibility effects are neglected), coherent vortices---points where the dynamical field $\psi(\mathbf{x},t)$ vanishes and the vorticity
is singular, see Sec.\ \ref{subsec:vortices} below.
	In this case, one can significantly reduce the complexity by tracking only the self-consistent motion of each vortex due to the flow induced by all the other vortices. Given the locations of the vortices $\{\mathbf{x}_j(t)\},$ with $j=1,\ldots N$, the $k$th vortex moves according to
\begin{equation}
	\label{eq:PV_EOM}
	\frac{\mathrm{d}\mathbf{x}_k(t)}{\mathrm{d}t}
	\equiv \dot{\mathbf{x}}_k(t)
	= \sum_{\substack{j=1 \\ j\neq k}}^N \frac{\kappa_j}{2\pi} 
		\frac{\hat{\mathbf{z}} \times \left( \mathbf{x}_k-\mathbf{x}_j \right)}{|\mathbf{x}_k - \mathbf{x}_j|^2},
\end{equation}
where $\hat{\mathbf{z}}$ points out of the plane, and $\kappa_j = 4\pi$ (in this paper we reserve the overdot notation for the total time derivative of a vortex position only).
 This is the point vortex (PV) model, which reduces the modelling problem from the full field PDE \eqref{eq:NLS} to $N$ 2-component ODEs \eqref{eq:PV_EOM}, where $N$ is the number of vortices. 

The PV model has enjoyed widespread use, particularly in the cold-atom community
\citep{ 
simula2014emergence,
billam2014onsager,
salman2016longrange,
reeves2017enstrophycascade, 
maestrini2019entropy,
lydon2022dipole},
but it is often motivated by the fact that the NLS equation can be recast into hydrodynamical form (see section \ref{subsec:Madelung}), and then appealing to the theorem that for incompressible inviscid flows, vorticity is transported along Lagrangian paths \citep{batchelor1967flluiddynamicsbook}. 
% and drawing on the long history of the PV model in hydrodynamics \citep{aref1983integrable, eyink2006onsager}, going back to its introduction by Helmholtz in 1858 \citep{helmholtz1858integrale}.
However, this vorticity transport theorem is only valid for well-behaved fields, whereas for the NLS equation the vorticity is singular at the vortex positions. The hydrodynamical description therefore fails precisely on the points where hydrodynamic intuition is invoked. 
It is therefore prudent to examine how, and under what conditions, the PV model can be derived rigorously from the NLS equation.

One approach to such a derivation, taken in \citep{neu1990vortices}, is to specify topological boundary conditions around the vortices, and solve for the motion of these boundaries in a way that is self-consistent with the dynamics of the rest of the field. 
	In this paper, we take an alternative approach. Here, we present a derivation of the
PV equation of motion from the Hamiltonian formulation of the NLS equation. In doing so, we distinguish between the parts of the Hamiltonian that lead to the PV equation of motion, and the parts that lead to the self-interaction energy of an NLS vortex. We numerically calculate the self-energy per vortex for the first time, to our knowledge.

This paper can be considered a companion piece to Ref.\ \citep{Bustamante2015BiotSavart}, in which the Biot-Savart model for line vortices was derived directly from the 3D NLS equation. The key difference between the present 2D case and the 3D case is that the Biot-Savart integral in the latter has a singularity which is regularised by means of a small-scale cutoff, whose value is determined accurately in Ref.\ \citep{Bustamante2015BiotSavart}.
	However, in the 2D problem the singularity is in fact integrable, and requires no cutoff. The derivation we present here is thus simpler than, and independent of, the 3D case.

Before commencing the derivation we briefly discuss two important areas of physics where Eq.\ \eqref{eq:NLS} is the governing equation, and where the PV model is frequently used as a reduced model of the dynamics.

\subsection{The 2D NLS equation in physical contexts}
\label{subsec:NLSphysical}

The 2D NLS equation is frequently used in low-temperature physics to model superfluid dynamics and turbulence in Bose-Einstein condensates of alkali gases in highly anisotropic 2D traps, and superfluid helium films
\citep{pitaevskiistringari2016book, 
nazarenko2006wave, 
nazarenko2014bose, 
billam2014onsager, 
stagg2015generation,
kwon2021sound}.
In this context, the NLS equation is more often referred to as the Gross-Pitaevskii equation \citep{gross1961structure, pitaevskii1961vortex}, and appears with physical units, as
\begin{equation}
	\label{eq:GPE}
	i\hbar\frac{\partial\psi}{\partial t} + \frac{\hbar^2}{2m}\nabla^2 \psi - g|\psi|^2\psi  = 0.
\end{equation}
In Eq.\ \eqref{eq:GPE}, $\psi$ represents the 1-particle wavefunction of the boson comprising the condensate or superfluid, $\hbar$ is the reduced Plank's constant, $m$ is the boson mass, and $g$ characterises the strength of interatomic $s$-wave interactions. 
	In the 2D case that we are concerned with here, Eq.\ \eqref{eq:GPE} is an effective equation in which the trapping potential in the $z$ direction confines the dynamics to the $x$-$y$ plane.  We treat this plane as being infinite, which is far from experimental reality, but is convenient for the theoretical derivation we present here. 

	Eq.\ \eqref{eq:GPE} can be written in terms of nondimensionalised variables, indicated by primes:	$\psi = \sqrt{\rho_0}\psi'$, where $\rho_0$ is the far-field number density of bosons in physical units, $t = (\hbar / g \rho_0) t'$, and $\mathbf{x} = \xi \mathbf{x}'$. Here $\xi = \hbar / \sqrt{2mg\rho_0}$ is the healing length in physical units, and is the lengthscale over which vortices recover to the background density $\rho_0$, see Sec.\ \ref{subsec:vortices}. 
	Finally, we move to a frame corotating with the condensate in the complex plane, via $\psi'(\mathbf{x}',t') = \psi''(\mathbf{x}',t') \exp(-it')$, i.e.\ the chemical potential $\mu=g\rho_0$ has been normalised to $1$. Dropping all primes immediately, we recover the nondimensionalised NLS Eq.\ \eqref{eq:NLS}.

Another principal application of Eq.\ \eqref{eq:NLS} is in nonlinear optics. 
	Here, the equation is the leading-order model for paraxial propagation of a linearly polarised, continuous wave laser beam through a homogeneous Kerr medium
\citep{dyachenko1992optical, 
carusotto2014superfluid, 
situ2020dynamics, 
eloy2021experimental}.
In this context, the equation appears with physical units as
\begin{equation}
	\label{eq:NLSoptics}
	2 i k_0 \frac{\partial\psi}{\partial z} + \nabla_\perp^2 \psi + 4 k_0^2 \frac{n_2}{n_0}  |\psi|^2  \psi = 0,
\end{equation}
and $\psi$ represents the complex envelope of the input electric field. The distance along the axis of propagation of the beam $z$ plays the role of a timelike variable, leaving the dynamics to take place in the 2D $x$-$y$ plane, as reflected in the perp symbol in the Laplacian. In Eq.\ \eqref{eq:NLSoptics} $k_0$ is the wavenumber of the laser in the medium, which has refractive index $n_0$, and $n_2$ is the Kerr coefficient. 

We make the transformation to nondimensional (primed) variables via $\psi = \sqrt{\rho_0}\psi'$, where $\rho_0$ is the far-field intensity in physical units, $z = (n_0 / 2 k_0 n_2 \rho_0) z'$, and $\mathbf{x} = \xi \mathbf{x}'$ with physical healing length $\xi = \sqrt{n_0 / 4 k_0^2 n_2 \rho_0}$. Transforming to the frame corotating with the condensate as above, and dropping the primes, we again recover Eq.\ \eqref{eq:NLS}.

For the rest of this paper we work in nondimensional units. In particular, the fiducial density $\rho_0$ and length $\xi$ both become $1$ in these units.

\section{Hamiltonian and hydrodynamic descriptions}

The NLS equation \eqref{eq:NLS} can be written in Hamiltonian form~\citep{ZLF1992book}
\begin{equation}
	\label{eq:HamEq}
	i\frac{\partial \psi}{\partial t} = \frac{\delta H}{\delta \psi^*},
\end{equation}
where the Hamiltonian functional
\begin{equation}
	\label{eq:Hpsi}
	H = \int \! \left[  \left| \nabla \psi \right|^2
					 +\frac{1}{2}\left( \left| \psi \right|^2-1\right)^2
				\right]  \mathrm{d}\mathbf{x}
\end{equation}
is equal to the energy of the system, and is conserved by evolution under Eqs.\ \eqref{eq:NLS}, \eqref{eq:HamEq}.
	We take the spatial domain to be $\mathbb{R}^2$.
	The normalisation of Sec.\ \ref{subsec:NLSphysical} give the far-field conditions $|\psi|^2 \to 1$ and $\nabla\psi\to0$  as $|\mathbf{x}|\to\infty$.
	The latter allows us to integrate by parts and neglect the boundary term at infinity.

\subsection{Hydrodynamic description and the Madelung transform}
\label{subsec:Madelung}

We can change the dynamical variable from the complex field $\psi(\mathbf{x},t)$ to the real fields $\rho(\mathbf{x},t)$ and $\phi(\mathbf{x},t)$ via the Madelung transform $\psi = \sqrt{\rho} \exp(i\phi)$ \citep{madelung1957begriffssystem, spiegel1980madelung}.
Substituting this into Eq.\ \eqref{eq:NLS}, and separating the real and imaginary parts, gives the equations
\begin{subequations}
	\label{eq:fluid}
	\begin{align}
		\frac{\partial \rho}{\partial t} + \nabla \cdot(\rho \mathbf{u}) &= 0 ,   \label{eq:fluid_conty} \\
		\frac{\partial \mathbf{u}}{\partial t} + (\mathbf{u}\cdot\nabla)\mathbf{u} 
			&= -\frac{\nabla\rho^2}{\rho} + \nabla\left(2\frac{\nabla^2\sqrt{\rho}}{\sqrt{\rho}}\right) . \label{eq:fluid_mom}
	\end{align}
\end{subequations}
We identify these as the mass and momentum conservation equations of an inviscid fluid with density $\rho(\mathbf{x},t)$ and velocity 
\begin{equation}
	\label{eq:velocity_phi}
	\mathbf{u}(\mathbf{x},t)=2\nabla\phi(\mathbf{x},t).
\end{equation} 
Thus we see that when $\rho$ and $\phi$ can be defined, the fluid description \eqref{eq:fluid} is equivalent to the original NLS equation \eqref{eq:NLS}. 
This fluid description has two contributions to the pressure on the RHS of Eq.\ \eqref{eq:fluid_mom}.
	The first is due to a polytropic equation of state $p=\rho^2$. The second due to the so-called quantum pressure, and represents the only difference between the fluid description of an NLS system and a real physical fluid.

Note that $\nabla\psi = [i (\nabla \phi) \sqrt{\rho} + \nabla \! \sqrt{\rho}]\exp(i\phi)$. Therefore the condition $\nabla\psi \to 0$ as $|\mathbf{x}|\to\infty$ implies that  in the far field both $|\mathbf{u}|\to0$ (by Eq.\ \eqref{eq:velocity_phi}, c.f.\ the slow-phase-variation condition considered in Ref.\ \citep{neu1990vortices}), and $\rho \to \text{const.}$, which we set to $1$.

\section{Quantised vortices and the Pitaevskii profile}
\label{subsec:vortices}
Even though the fluid velocity $\mathbf{u}$ obtained from the Madelung transform is manifestly irrotational, vortices may still appear in the system. These are the points $\{\mathbf{x}_j\}$ where the density $\rho$, and hence $\psi$, vanish. Consequently, the phase $\phi$, and hence the velocity $\mathbf{u}$, is undefined at these points. The Madelung transformation cannot be made there. However, it is precisely at these points where we wish to apply hydrodynamical intuition. 
It is this paradox that motivates the present detailed derivation of the PV model from the 2D NLS equation.

In order to see that the phase defect points represent vortices, consider any closed contour $C$ that embraces such a defect. On traversing $C$, the phase $\phi$ must change by a multiple of $2\pi$ in order to keep $\psi$ single-valued. One can then define the fluid circulation around $C$,
\begin{equation}
	\label{eq:circulation}
	\kappa = \oint_C\!  \mathbf{u}  \cdot  \mathrm{d}\mathbf{l}
				  = \oint_C\!  2\nabla\phi \cdot  \mathrm{d}\mathbf{l}
				  = 2 [\phi]_C
				  = \pm 4\pi n, 
\end{equation}
where $n$ is a positive integer. By contrast, $\kappa=0$ on contours that embrace no phase defects. We thus conclude that the phase defects are vortices, and note that the circulation of each such vortex is quantised in integer multiples of $4\pi$. However, vortices with $n\geq 2$ are unstable and decay into elementary vortices with $n=1$ on any general smooth change of the field $\phi$ (see Ref.\ \citep{patrick2022origin} for the $n=2$ case, and references therein).
	Therefore, we only consider elementary point vortices with $\kappa=\pm 4\pi$ for the remainder of this work. 

Considering an isolated elementary vortex at $\mathbf{x}=0$ and taking a circular contour with radius $r=|\mathbf{x}|$, Eq.\ \eqref{eq:circulation} gives the vortex velocity profile  
\begin{equation}
	\label{eq:PitaevskiiVortex_u}
	\mathbf{u}(\mathbf{x})=\frac{2}{r}\hat{\bm{\theta}}(\mathbf{x}),
\end{equation}
where $\hat{\bm{\theta}}(\mathbf{x})$ is an azimuthal unit vector. 

The density profile of an isolated NLS vortex was found by Pitaevskii \citep{pitaevskii1961vortex} (although Ginzburg and Pitaevskii had earlier found the same vortex solution when examining superfluid helium in the framework of Landau's theory of phase transitions \citep{ginzburg1958theory}).
	Setting $\phi(r,\theta)=\theta$ for an elementary vortex, we assume a time-independent solution with a radially-symmetric density profile: $\psi_v(\mathbf{x})=R(r) \exp(i\theta)$. Substituting this into Eq.\ \eqref{eq:NLS} gives the ODE
\begin{equation}
	\label{eq:PitaevskiiVortex}
		\frac{\mathrm{d}^2 R}{\mathrm{d}r^2}  
	+  \frac{1}{r}\frac{\mathrm{d}R}{\mathrm{d}r}  
	-  \frac{1}{r^2} R  
	+  (1-R^2) R  
	=  0,
\end{equation}
with the boundary conditions $R(0)=0$ and $R(r)\to1$ as $r\to\infty$.
We refer to $R(r)$ as the Pitaevskii profile, and the associated complex field $\psi_v(\mathbf{x})$ as the Pitaevskii vortex solution.

The asymptotics of $R(r) = \sqrt{\rho(r)}$ can be found by balancing dominant terms in Eq.\  \eqref{eq:PitaevskiiVortex}.
Deep in the vortex core we balance the second and third terms and obtain a linear profile which we write as 
\begin{equation*}
	R(r)  \to \sqrt{2}\alpha r \quad \mathrm{as} \quad r\to0,
\end{equation*}
with $\alpha=\mathrm{const}$.
Far from the vortex we balance the first and fourth terms of Eq.\  \eqref{eq:PitaevskiiVortex}. Noting that $|1-R(r)| \ll 1$, we obtain exponential convergence to the asymptotic value $\sqrt{\rho_0}=1$. 
More generally, we solve Eq.\  \eqref{eq:PitaevskiiVortex} via the highly-accurate numerical method of Ref.\ \cite{Bustamante2015BiotSavart}. 
This method improves on other less accurate methods of calculating the Pitaevskii profile, such as the Pad\'e approximation method \citep{berloff2004pade} (see Ref.\ \citep{Bustamante2015BiotSavart} for relevant comparisons).
 
\begin{figure}
     \includegraphics[width=0.99\columnwidth]{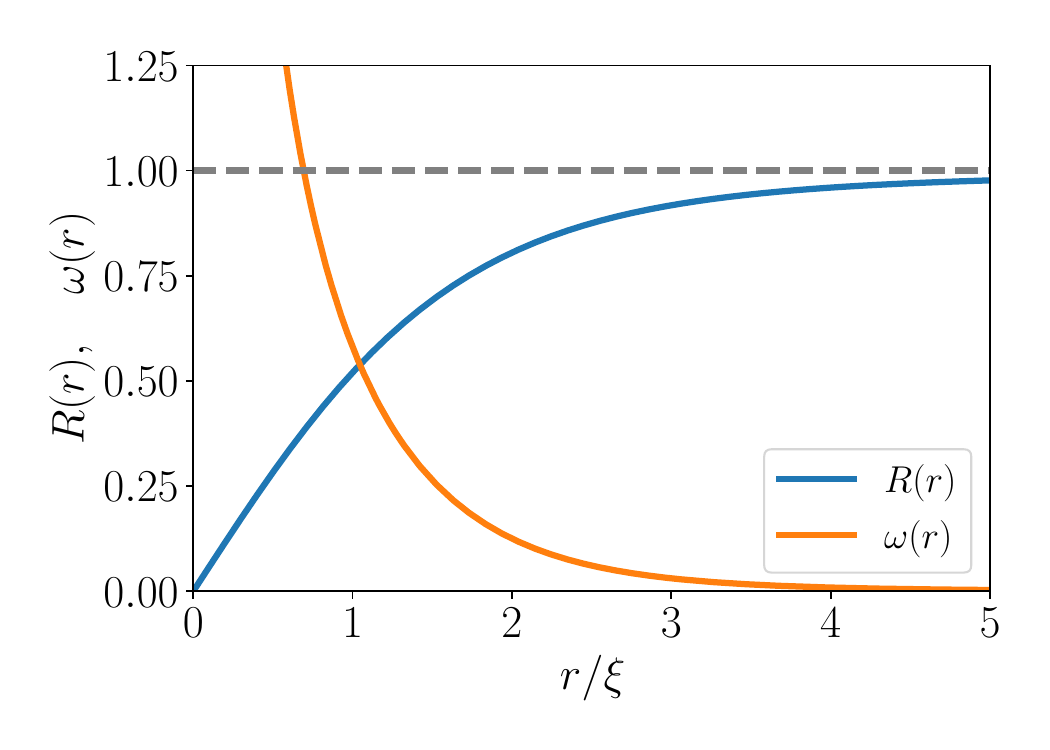}
    \caption{
    \label{fig:Romegaplot}
    Profiles of $R(r) = \sqrt{\rho(r)}$ (solid blue), and the pseudovorticity field $\omega(r)$ (solid orange) of an isolated Pitaevskii vortex. 
    The horizontal grey dashed line represents the uniform condensate.
    }
\end{figure}

We plot the vortex profile $R(r)$ in Fig.\ \ref{fig:Romegaplot}.
	The characteristic radius over which the vortex heals to $\rho_0$ is the healing length $\xi \! \sim \! 1/\sqrt{\rho_0}$, found by balancing the nonlinear and linear terms in Eq.\  \eqref{eq:NLS}.

The velocity profile \eqref{eq:PitaevskiiVortex_u}, and the density profile obtained from Eq.\ \eqref{eq:PitaevskiiVortex}, are fixed for all isolated elementary vortices in an NLS system (as opposed the arbitrary profiles of vortices in classical fluids). 
For an ensemble of well-separated vortices, these profiles will become asymptotically correct as $r\to 0$.

\section{Assumptions of this derivation}
\label{sec:assumptions}
As well as the assumption of a spatially constant condensate in the far field, in this work we restrict ourselves to the case where we have $N$ elementary vortices located at positions $\mathbf{x}_j(t)$, $1\leq j\leq N$, where $N$ is finite.
	We take the number of vortices with positive and negative circulation to be equal, which we term a neutral ensemble. This is to ensure that in the far field, the overall anti-clockwise rotation induced by the positive vortices exactly balances the overall clockwise rotation induced by the negative vortices. In other words, the system has no net angular momentum, as required by the condtion $|\mathbf{u}|\to0$ at infinity.
 
We further assume that the vortices start well-separated, and remain so during the dynamics, i.e.\ for every pair of vortices labelled by $j$ and $m$, the inter-vortex distance $\ell_{jm}(t) = |\mathbf{x}_j(t) - \mathbf{x}_m(t)|$ is always much greater than the healing length $\xi$.
	In addition, we assume the flow $\mathbf{u}(t)$ to be incompressible, i.e.\ all motions 
of the vortices remain subsonic ($\dot{\mathbf{x}}_k(t) \ll 1$), and that there are no significant acoustic excitations present in the intial field.
	This assumption is there to retain consistency with the PV model that will be the outcome of this derivation, and which does not describe compressible dynamics.

While these assumptions might not be the most general in relation to physically-realisable scenarios, we believe they are the minimal set of assumptions that enable us to derive the PV model from the NLS equation with the degree of rigour that we seek to employ.

\section{Transforming the Hamiltonian}
\label{sec:transformingH}

After the Madelung transformation, the Hamiltonian \eqref{eq:Hpsi} becomes
$H = H_K + H_0$, where 
\begin{equation}
	\label{eq:H_K}
	H_K = \frac{1}{2} \int \!  \frac{ \rho |\mathbf{u}|^2 }{2} \,\mathrm{d}\mathbf{x}
\end{equation}
is the kinetic energy of the fluid that is described by Eqs.\ \eqref{eq:fluid}.
	The spatial regions that contribute to $H_K$ are delocalised due to the slow $\sim\!1/r$ decay of velocity with distance $r$ from the vortex cores. On the other hand,
\begin{equation}
	\label{eq:H_0}
	H_0 = \frac{1}{2} \int \! \left[ (\rho-1)^2 + 2 \left| \nabla \sqrt{\rho} \right|^2 \right] \,\mathrm{d}\mathbf{x}
\end{equation}
represents the total internal energy derived from the hydrodynamic and quantum pressures in Eq.\ \eqref{eq:fluid_mom}.

Despite the singularity of the Madelung transform at the vortex positions $\{\mathbf{x}_j\}$,  the integrals in Eqs.\ \eqref{eq:H_K} and \eqref{eq:H_0} are well-defined over all of $\mathbb{R}^2$,  since $\rho$ is well-defined everywhere, and zero at the vortex positions. 
	In particular Eq.\  \eqref{eq:H_K} picks up zero contribution at the vortex positions, as we show in Sec.\ \ref{subsubsec:rewritingHK}.

\subsubsection{Calculation of the internal energy per vortex}

The integrand in $H_0$ is only significant when the density deviates from the background value of $1$, so the contributions to $H_0$ are localised to the vortex cores. As the cores are well-separated by assumption, we have $N$ identical contributions, and can speak of the internal energy per vortex
\begin{align}
	\frac{H_0}{N} & = \pi \bigintssss_0^\infty \! \left[ (R^2-1)^2 + 
								2 \left( \frac{\mathrm{d} R}{\mathrm{d}r} \right)^2 \right] r\, \mathrm{d}r  \nonumber \\
						&= 4.8951725778 . 	\label{eq:H_0_result}
\end{align} 
(This result was obtained numerically in Ref.\ \citep{Bustamante2015BiotSavart}).

\subsubsection{Rewriting the kinetic energy}
\label{subsubsec:rewritingHK}

Following Ref.\ \citep{Bustamante2015BiotSavart}, we wish to write $H_K$ in terms of new flow variables that have constant density to leading order, and that are regular at the vortex positions. 
	We therefore introduce a new field $\mathbf{v}(\mathbf{x},t)$, which we term the pseudovelocity:
\begin{equation}
	\label{eq:v_field}
	\mathbf{v} = \sqrt{\frac{\rho}{2}} \mathbf{u} = \nabla \times \left(\Psi \hat{\mathbf{z}}\right) .
\end{equation}
We have expressed the pseudovelocity in terms of the streamfunction $\Psi(\mathbf{x},t)$ as the flow is incompressible by assumption.
The corresponding pseudovorticity field is
$
	\omega = \left(\nabla \times \mathbf{v} \right)\cdot\hat{\mathbf{z}} = - \nabla^2 \Psi 
$, which has the formal solution
\begin{equation}
	\label{eq:Psi_field}
	\Psi(\mathbf{x},t) = - \int\!  G(|\mathbf{x}-\mathbf{x}'|) \, \omega(\mathbf{x}',t) \, \mathrm{d}\mathbf{x}' ,
\end{equation}
where, for the infinite 2D domain, the Green's function is $G(|\mathbf{x}-\mathbf{x}'|) = (2\pi)^{-1} \log(|\mathbf{x}-\mathbf{x}'|)$. We will shortly comment on the regularity of the integral in Eq.\ \eqref{eq:Psi_field}
 in our case.
%the Green's function satisfies $\nabla^2 G(|\mathbf{x}-\mathbf{x}'|) = \delta(\mathbf{x}-\mathbf{x'})$;

For an isolated vortex at the origin, the profile of the $\mathbf{v}$ field is 
\begin{equation*}
%	\label{eq:vprofile}
	\mathbf{v} = \frac{\sqrt{2\rho(r)}}{r}\hat{\bm{\theta}} ,
\end{equation*}
with $v = |\mathbf{v}|\to 2\alpha$ as $r\to 0$.
The corresponding $\omega$ profile is
\begin{equation}
	\label{eq:omegaprofile}
	\omega = \frac{1}{r}\frac{\partial}{\partial r}(r v) \to \frac{2\alpha}{r} 
	\quad \mathrm{as} \quad 
	r\to 0.
\end{equation}
Recalling the behaviour of $R(r)=\sqrt{\rho(r)}$, we conclude that the $\omega(r)$ profile is a strong spike with characteristic lengthscale $\xi$, and decays  to zero rapidly outside the vortex core. This is shown in Fig.\ \ref{fig:Romegaplot}.

In terms of the $\mathbf{v}$ field, the kinetic energy becomes
\begin{equation}
	\label{eq:H_Psiomega}
	%H_K  = \frac{1}{2} \int\! v(\mathbf{x},t)^2 \, \mathrm{d}\mathbf{x} \\
	H_K  = \frac{1}{2} \int\! \left|{\bf v}(\mathbf{x},t)\right|^2 \, \mathrm{d}\mathbf{x}
		= \frac{1}{2} \int\! \Psi(\mathbf{x},t)\ \omega(\mathbf{x},t) \, \mathrm{d}\mathbf{x},
\end{equation}
where we have used Eq.\ \eqref{eq:v_field}, and integrated by parts. 

The latter step requires some care as $\omega$ is singular at the vortex locations $\{\mathbf{x}_j\}$. 
	Therefore, to carry out the integration by parts, we consider the following limiting procedure. 
	We first consider taking the integral over the perforated domain $\mathbb{R}^2 \setminus \cup_{j=1}^N \mathcal{P}_j$, i.e.\ we remove patches of finite area around each vortex, with $\mathcal{P}_j$ the patch around the $j$th vortex.
	Each patch is constructed so that its boundary is a closed streamline of $\mathbf{v}$. As it is a streamline, $\Psi$ is constant on the boundary. The boundary around the $j$th patch, $\partial \mathcal{P}_j$, contributes a term to Eq.\ \eqref{eq:H_Psiomega} of
\begin{equation*}
	\frac{1}{2}\Psi \oint_{\partial \mathcal{P}_j} \mathbf{v}\cdot\mathrm{d}\mathbf{l} ,
\end{equation*}
with the integral around $\partial \mathcal{P}_j$ taken clockwise.
	Reversing the integration direction, and using Stokes' theorem to relate this to an integral over the interior of $\mathcal{P}_j$, the $j$th boundary term becomes $- (\Psi/2) \int_{\mathcal{P}_j} \omega \,\mathrm{d}\mathbf{x}$. Next, we take any sequence of progressively smaller patches localised on the vortices. As the area of each patch progressively shrinks, the bounding streamlines become more and more circular with radius $r$, so that we can use Eq.\ \eqref{eq:omegaprofile} and write the $j$th boundary term as
\begin{equation*}
	 - \frac{1}{2}\Psi \int_{\mathcal{P}_j} \frac{2\alpha}{r'} \, r' \mathrm{d}r' \, \mathrm{d}\theta  \,\to\, 0 
	 \quad \mathrm{as} \quad 
	r\to 0.
\end{equation*}
Thus we restore the domain of integration in Eq.\ \eqref{eq:H_Psiomega} to $\mathbb{R}^2$. 

Using Eq.\ \eqref{eq:Psi_field}, the kinetic energy becomes
\begin{equation}
	\label{eq:HomegaG}
		H_K = -\frac{1}{2} \int\! \omega(\mathbf{x},t)\, \omega(\mathbf{x}',t)\, G(|\mathbf{x} - \mathbf{x}'|) \, \mathrm{d}\mathbf{x}\,\mathrm{d} \mathbf{x}'.
\end{equation}

\subsection{Dividing the kinetic energy into local and distant contributions}

\begin{figure*}
     \includegraphics[width=0.99\textwidth]{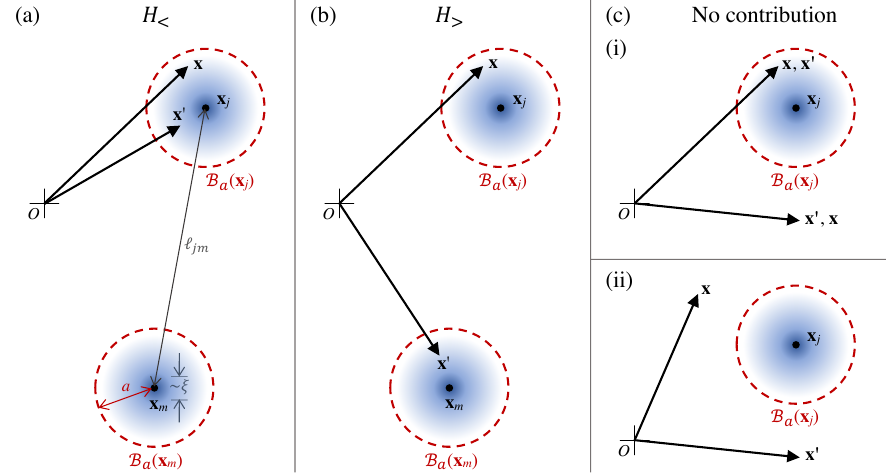}
    \caption{
    \label{fig:Hcontribs}
    Typical contributions to $H_K$. 
    (a) Contribution to $H_<$ when $\mathbf{x}, \mathbf{x}'$ lie within the same ball $\mathcal{B}_a(\mathbf{x}_j)$ for some vortex $j$. 
    		(We also indicate the scale separation $\xi \ll a \ll \ell=\min(\ell_{jm})$.)
    (b) Contribution to $H_>$ when $\mathbf{x}\in\mathcal{B}_a(\mathbf{x}_j)$ and $\mathbf{x}'\in\mathcal{B}_a(\mathbf{x}_m)$ for different vortices $j,m$.
    (c) No contribution when (i) one, or (ii) both of $\mathbf{x}, \mathbf{x}'$ do not lie inside $\mathcal{B}_a(\mathbf{x}_j)$ for any vortex $j$.
    }
\end{figure*}

The next step is to separate out the contributions to $H_K$ that are local to one vortex from those that involve spatially distant parts of the domain. 
We do this by introducing an intermediate lengthscale $a$ such that $\xi \ll a \ll \ell=\min(\ell_{jm})$, and then writing
$H_K = H_< + H_> $, which we define below.

\subsubsection{Self-interaction kinetic energy $H_<$}

$H_<$ contains the sum of all contributions where both integration variables $\mathbf{x}$ and $\mathbf{x}'$ lie within distance $a$ of the same vortex, say the $j$th, i.e.\ within the ball $\mathcal{B}_a(\mathbf{x}_j)$, as shown in Fig.\ \ref{fig:Hcontribs}(a).

Clearly, $H_<$ gives the total energy due to the self-interaction of the $N$ vortices. 
	By assumption, the vortices of our system are well-separated; to leading order we treat each as giving an identical contribution. Therefore,
\begin{equation}
	\label{eq:Hless}
	H_< = -\frac{N}{4\pi} \!\!\!\!\!\! \bigintssss\displaylimits_{\mathbf{x},\mathbf{x}' \in \mathcal{B}_a(\mathbf{x}_j)} \!\!\!\!\!\!
						\omega(\mathbf{x},t)\omega(\mathbf{x}',t) \log(|\mathbf{x}-\mathbf{x}'|) \, \mathrm{d}\mathbf{x}\,\mathrm{d}\mathbf{x}'.
\end{equation}
Furthermore, the $\xi \ll a$ assumption allows us to formally send $a\to\infty$ when calculating this integral. 

We change coordinates from $(\mathbf{x},\mathbf{x}')$ to polar coordinates $(r,\theta, r',\varphi)$, where $\varphi$ is the angle between $\mathbf{x}$ and $\mathbf{x}'$. 
Using Eq.\ \eqref{eq:omegaprofile} for the $\omega(r)$ profile and integrating over $\theta$ immediately, we obtain
\begin{equation}
	\begin{aligned}
		\label{eq:Hless_working}
		\frac{H_<}{N}
				= & - \int_0^\infty \int_0^\infty \int_0^{2\pi} 
						\frac{\mathrm{d} R(r)}{\mathrm{d} r} \frac{\mathrm{d} R(r')}{\mathrm{d} r'} 
				\\
					&\times	\log\left(\sqrt{r^2 + r'^2 - 2 r r' \cos\varphi}\right) 
					\mathrm{d}\varphi \,  \mathrm{d}r' \, \mathrm{d}r .
	\end{aligned}
\end{equation}
The $\varphi$ integration can be done analytically as follows. We can use $\cos(2\pi-\varphi)=\cos\varphi$ to double the integrand while halving the integration domain. Then, assuming first that $\beta = r/r' \geq 1$, we write
\begin{align*}
	I &= \int_0^\pi \! \log\left(r^2 + r'^2 - 2 r r' \cos\varphi \right) \, \mathrm{d}\varphi \\
	&= \int_0^\pi \! \left[ \log(r'^2) + \log\left( 1 - 2 \beta \cos\varphi + \beta^2 \right) \right] \mathrm{d}\varphi \\
	&= \pi \log(r'^2) + \pi \log( \beta^2 ) \\
	&= \pi \log(r^2),
\end{align*}
where in the penultimate step we have used the Leibniz integral rule \citep{woods1926advcalc}.
Likewise, assuming $\beta = r'/r \geq 1$ gives $I=\pi \log(r'^2)$.
Combining the two results, we obtain $I = \pi \log\left[ \max(r^2,r'^2) \right]$. 

We substitute this into Eq.\ \eqref{eq:Hless_working}, and integrate over $r$ and $r'$ numerically using the method given in Ref.\ \citep{Bustamante2015BiotSavart} for $\mathrm{d}R(r)/\mathrm{d}r$, obtaining
\begin{align}
	\frac{H_<}{N}  &=  -  \pi \int_0^\infty \!\! \int_0^\infty \! 
										\frac{\mathrm{d} R(r)}{\mathrm{d} r} \frac{\mathrm{d} R(r')}{\mathrm{d} r'} 
										\log \left[\max(r^2,r'^2)\right]  \mathrm{d}r  \, \mathrm{d}r' \nonumber  \\
					 	 &=  -  2.5020210587. \label{eq:Hless_result}
\end{align}

\subsubsection{Total self-energy per vortex}

Using Eqs.\ \eqref{eq:H_0_result} and \eqref{eq:Hless_result}, we can now write down the total self-energy per vortex:
\begin{equation}
	\frac{H_\mathrm{self}}{N} = \frac{H_0 + H_<}{N} 
										  = 2.3931515191. 	\label{eq:Hself_result}
\end{equation}
Note that $H_\mathrm{self}/{N}$ is positive despite $H_</N$ being negative. As discussed in Sec.\ \ref{subsubsec:wellsep_incompr}, this energy contributes to the energy of acoustic waves that are generated when vortices are allowed to annihilate.

To recap, Eq.\ \eqref{eq:Hself_result} gives the energy that is localised to a vortex core, in a neutral ensemble of vortices, situated in an infinite domain (and in accordance with the other assumptions stated in Sec.\ \ref{sec:assumptions}). 
	Other authors give different expressions for the energy of a point vortex. For example, Ginzburg and Pitaevskii \citep{pitaevskii1961vortex, ginzburg1958theory} calculate the total energy (corresponding to our total Hamiltonian $H$) of a single isolated vortex in the centre of a disc, with a large-scale cutoff associated with the disc radius $R$. By contrast, here we are concerned with the neutral $N$-vortex configuration in an infinite domain, and have divided the contributions to the energy into $H_\mathrm{self}$ and $H_>$. As we shall see, it is $H_>$ that leads to the mutual interaction between vortices, and hence the PV equation of motion \eqref{eq:PV_EOM}.	
	 Maestrini \citep{maestriniPhDthesis} carries out a calculation that is similar to that of this section, but he discards the terms that lead to $H_\mathrm{self}$. In addition, his calculation relates to an $N$-vortex ensemble that is not necessarily neutral. Any imbalance between positive and negative vortices results in a contribution to the energy associated with the cutoff at $R$, just as in Ginzburg and Pitaevskii's case. We discuss this contribution in Sec.\ \ref{subsubsec:neutral_ensemble}.

\subsubsection{Mutual-interaction kinetic energy $H_>$}

$H_>$ contains all cases where $\mathbf{x}, \mathbf{x}'$ do not lie within the same ball $\mathcal{B}_a(\mathbf{x}_j)$. Now recall that $\omega(\mathbf{x},t)$ is sharply peaked over the vortex cores, and decays rapidly to zero outside them, over a lengthscale $\sim \! \xi$.
	If either $\mathbf{x}$ or $\mathbf{x}'$ lie outside any vortex core, in particular outside any ball $\mathcal{B}_a(\mathbf{x}_j)$, then the corresponding $\omega$ will be vanishingly small, and there will be practically no contribution to $H_>$. This is illustrated in Fig.\  \ref{fig:Hcontribs}(c).
	Therefore $H_>$ only picks up contributions where $\mathbf{x}$ and $\mathbf{x}'$ lie within distance $a$ of different vortices, see Fig.\  \ref{fig:Hcontribs}(b)
(if they both lie within $a$ of the same vortex, the contribution is to $H_<$). We therefore have
\begin{equation}
	\label{eq:Hgreater_int}
	H_> = - \sum\displaylimits_{\substack{
															j,m=1 \\
															j\neq m }
															}^N
				\frac{1}{4\pi} 
									 \!\!\!\!\!\! \bigintssss
													\displaylimits_{\substack{
																					\mathbf{x} \in \mathcal{B}_a(\mathbf{x}_j),  \\
																					\mathbf{x}' \in \mathcal{B}_a(\mathbf{x}_m)}
																		}
													\!\!\!\!\!\!
													\omega(\mathbf{x},t)\omega(\mathbf{x}',t) 
													\log(|\mathbf{x}-\mathbf{x}'|) \, \mathrm{d}\mathbf{x}\,\mathrm{d}\mathbf{x}'.
\end{equation}
Clearly this contribution to the Hamiltonian reflects the mutual interactions of each vortex with every other vortex.
	
Given the restriction of $\mathbf{x},\mathbf{x}'$ to different vortex neighbourhoods, the integral only has contributions when $|\mathbf{x}-\mathbf{x}'|\gtrsim \ell$. 
On these scales, $\log(|\mathbf{x}-\mathbf{x}'|)$ is a slowly-varying function compared to $\omega(\mathbf{x},t)$, which is peaked sharply around $\{\mathbf{x}_j(t)\}$.
	Thus, we can treat the logarithm as a constant over the width of each vortex core, i.e.\ under the integral in Eq.\ \eqref{eq:Hgreater_int} we have
$\omega(\mathbf{x},t) \approx (\kappa_j/\sqrt{2})  \delta\left(\mathbf{x}-\mathbf{x}_j(t)\right)$, 
%\begin{equation*}
%		\omega(\mathbf{x},t) \approx \frac{\kappa_j}{\sqrt{2}} \, \delta\left(\mathbf{x}-\mathbf{x}_j(t)\right),
%\end{equation*}
where $\kappa_j = \pm 4\pi$ is the elementary quantum of circulation of each vortex $j$. Using the properties of the Dirac delta, we can therefore write
\begin{equation}
	\label{eq:Hgreater}
	H_> = -\frac{1}{8\pi}  \sum
										\displaylimits_{\substack{
																	j,m=1 \\
																	j\neq m }
															}
															^N 
															\kappa_j\kappa_m \,
															\log(|\mathbf{x}_j-\mathbf{x}_m|).
\end{equation}

\section{From the Hamiltonian to the point vortex model}

Having obtained the expressions for $H_<$ and $H_>$, we now return to the original NLS in Hamiltonian form \eqref{eq:HamEq}.
	To turn this into an equation of motion for a particular vortex $k$, we must change the dynamical variables from the fields $\psi, \psi^*$ to the vortex position coordinates $\mathbf{x}_k$. We do this by multiplying Eq.\ \eqref{eq:HamEq} by $\partial\psi^*/\partial\mathbf{x}_k$, adding the resulting complex conjugate ($c.c.$), and integrating the result over the $\mathbf{x}$-plane:
\begin{equation}
\label{eq:dHdxk}
	\begin{aligned}
		&i \int \frac{\partial\psi^*}{\partial\mathbf{x}_k}  \frac{\partial\psi}{\partial t} d\mathbf{x}    \,+\,   c.c. \\
		&= 			i \int \left[  \frac{\delta H}{\delta \psi^*}\frac{\partial \psi^*}{\partial \mathbf{x}_k} 
						+  \frac{\delta H}{\delta \psi}\frac{\partial \psi}{\partial \mathbf{x}_k}  \right] d\mathbf{x}
		\equiv 	\frac{ \partial H}{\partial \mathbf{x}_k} .
	\end{aligned}
\end{equation}

We now work on the LHS of Eq.\ \eqref{eq:dHdxk}. Recalling the Pitaevskii vortex solution $\psi_v$, we can construct the field $\psi(\mathbf{x},t)$ for $N$ well-separated vortices as the product of $N$ displaced Pitaevskii vortices:
\begin{equation}
	\label{eq:psiprod}
	\psi(\mathbf{x},t) =    \prod_{j=1}^N \psi_v(\mathbf{x}-\mathbf{x}_j(t)) =  \prod_{j=1}^N e^{i s_j\theta_j} R_j,
\end{equation}
where $R_j = R(|\mathbf{x} - \mathbf{x}_j|)$ the Pitaevskii profile of the $j$th vortex, $s_j = \kappa_j/4\pi = \pm 1$ is its sign,  and $\theta_j$ is the polar angle around it:
\begin{equation*}
	\theta_j (\mathbf{x}-\mathbf{x}_j) = \arctan\!\left( \frac{y-y_j}{x-x_j} \right).
\end{equation*}
Note that the time dependence in $\psi(\mathbf{x},t)$ comes via the  $\mathbf{x}_j (t)$'s only, due to the assumptions of well-separated vortices and incompressibility, which ensure that the Pitaevskii profiles remain rigid as the vortices move. This implies that
\begin{equation*}
	\frac{\partial\psi}{\partial t} = \sum_{j=1}^N \frac{\partial\psi}{\partial\mathbf{x}_j} \cdot \dot{\mathbf{x}}_j .
\end{equation*}
Equation \eqref{eq:psiprod} implies
\begin{equation*}
	\frac{\partial\psi}{\partial\mathbf{x}_j} 
		= \left(\prod \displaylimits_{\substack{
																	m=1 \\
																	m\neq j }
															}
															^N 
				R_m e^{i s_m \theta_m} \right) 
			\frac{\partial}{\partial \mathbf{x}_j} \! \left(R_j e^{i s_j \theta_j}\right) ,
\end{equation*}
giving for the LHS of Eq.\ \eqref{eq:dHdxk},
\begin{multline}
\label{eq:LHS_dHdxk}
		 \bigintsss \!  \sum_{j=1}^N   \left( \prod \displaylimits_{\substack{
																	m=1 \\
																	m\neq j,k }
															}
															^N 
													R_m^2 	\right) 		  
				 \Biggl\{ i
				\left[ R_j R_k e^{i(s_k\theta_k - s_j\theta_j)}  \right]^{ (1-\delta_{jk}) } \\
						\times \, \left[ \dot{\mathbf{x}}_j \cdot  \frac{\partial}{\partial\mathbf{x}_j} \!\left( R_j e^{is_j\theta_j} \right) \right] 
 							   \left[ \frac{\partial}{\partial\mathbf{x}_k} \!\left( R_k e^{-is_k\theta_k} \right) \right] 		
 				 \Biggr\}
 					 \mathrm{d}\mathbf{x} + c.c.
\end{multline}
where $\delta_{jk}$ is a Kronecker delta. 

In Eq.\ \eqref{eq:LHS_dHdxk} we have enclosed the complex factors in the summand by large braces. For brevity, we extract them for further manipulation. Using the complex conjugate, the factor in braces becomes the real vector quantity
\begin{equation}
	\label{eq:vectorpart_dHdxk}
%	\begin{split}
%		&i \left[e^{i(s_k\theta_k - s_j\theta_j)}  \right]^{ (1-\delta_{jk}) }
%				\left[ \dot{\mathbf{x}}_j \cdot  \frac{\partial}{\partial\mathbf{x}_j} \!\left( R_j e^{is_j\theta_j} \right) \right] \\
% 			& \qquad \qquad \times \left[ \frac{\partial}{\partial\mathbf{x}_k} \!\left( R_k e^{-is_k\theta_k} \right) \right]  + c.c.  \\ 
%		&= i \left\{e^{i(s_k\theta_k - s_j\theta_j)}  \right\}^{ (1-\delta_{jk}) }  
%					\left[ \dot{\mathbf{x}}_j \!\cdot\!  \left( \frac{\partial  R_j}{\partial\mathbf{x}_j}  
%					+ is_j\frac{\partial\theta_j}{\partial\mathbf{x}_j}  R_j \right)  e^{is_j\theta_j} \right]  			
%					\left[ \left(\frac{\partial R_k}{\partial\mathbf{x}_k}  -is_k \frac{\partial\theta_k}{\partial\mathbf{x}_k}  R_k \right)e^{-is_k\theta_k} \right] 
%			+ c.c. \nonumber \\
%		&= 
		2 \left[
					s_k R_k \left(\dot{\mathbf{x}}_j \!\cdot\! \frac{\partial R_j}{\partial \mathbf{x}_j}\right) \frac{\partial \theta_k}{\partial\mathbf{x}_k} 
				   -s_j R_j \left(\dot{\mathbf{x}}_j \!\cdot\! \frac{\partial \theta_j}{\partial \mathbf{x}_j}\right) \frac{\partial R_k}{\partial\mathbf{x}_k}
				\right].
%		\end{split}
\end{equation} 

We now define the radial vector from the $j$th vortex $\mathbf{r}_j = \mathbf{x}-\mathbf{x}_j$, (with corresponding length $r_j$), and note that $\partial_{\mathbf{x}_j}\theta_j = -\hat{\mathbf{z}} \times \mathbf{r}_j/(r_j)^2$, while  $\partial_{\mathbf{x}_j}R_j =  -R_j' \mathbf{r}_j/r_j$. 
%where $R_j'\equiv \partial_{r_j} R_j(r_j)$.

Recall that we are considering the motion of vortex $k$. Taking into account that $\partial_{\mathbf{x}_j}\theta_j$ decays as $1/r_j$, and that $R_j$ heals exponentially to 1 (and $R_j'$ decays faster than exponentially to 0) on the length scale $\sim\!\xi\ \! \ll \! \ell_{jk}$ if $j\neq k$, we see that the main contribution to the sum in \eqref{eq:LHS_dHdxk} comes from the $j=k$ term. Retaining only this term, Eq.\ \eqref{eq:vectorpart_dHdxk} becomes
\begin{align*}
	&2 \frac{s_k R_k R_k'}{r_k^3}\left[
			 (\dot{\mathbf{x}}_k \cdot \mathbf{r}_k)  \, (\hat{\mathbf{z}} \times \mathbf{r}_k)
		   - \bigl( \dot{\mathbf{x}}_k \cdot (\hat{\mathbf{z}} \times \mathbf{r}_k) \bigr) \,\mathbf{r}_k
		\right] \\
	&= 2 \frac{s_k R_k R_k'}{r_k^3}\left[
			\dot{\mathbf{x}}_k \times \bigl(\left( \hat{\mathbf{z}} \times \mathbf{r}_k\right) \times \mathbf{r}_k \bigr)
		\right] \\
		&= 2 \frac{s_k R_k R_k'}{r_k}   \left(  \hat{\mathbf{z}}  \times \dot{\mathbf{x}}_k  \right),
\end{align*}
where we have used the vector triple product formula %$\mathbf{a}\times (\mathbf{b}\times\mathbf{c}) = (\mathbf{a}\cdot\mathbf{c}) \mathbf{b} - (\mathbf{a}\cdot\mathbf{b})\mathbf{c}$
 and the fact that $\hat{\mathbf{z}} \perp \mathbf{r}_k$.

Therefore, for the LHS of Eq.\ \eqref{eq:dHdxk} we have
\begin{align*}
	\bigintsss \! &  \left( \prod \displaylimits_{\substack{
																	m=1 \\
																	m\neq k }
															}
															^N 
			R_m^2 \right) 
			2 \frac{s_k R_k R_k'}{r_k}   \left(  \hat{\mathbf{z}}  \times \dot{\mathbf{x}}_k  \right) \mathrm{d}\mathbf{x} \\
	&\approx
		2 s_k  \left(  \hat{\mathbf{z}}  \times \dot{\mathbf{x}}_k  \right)  \int_0^\infty \frac{ R_k R_k'}{r_k}  \, 2\pi \, r_k \, \mathrm{d}r_k \\
	&= 2\pi s_k \left(  \hat{\mathbf{z}}  \times \dot{\mathbf{x}}_k  \right) ,
\end{align*}
where we have again used the fact that the vortices are well-separated, so $R_m^2\approx 1$ in the vicinity of $\mathbf{x}_k$ if $m\neq k$, and also used the asymptotics of $R_k$.

Finally, Eq.\ \eqref{eq:dHdxk} becomes
\begin{equation}
	\label{eq:dHdxk_penultimate}
	\frac{\kappa_k}{2} \left( \dot{\mathbf{x}}_k \times \hat{\mathbf{z}} \right) = -\frac{\partial H}{\partial \mathbf{x}_k}.
\end{equation}
For the RHS we note that $H = H_0 + H_< + H_>$, and that $H_0$ and $H_<$ are constants that vanish when differentiated. We recover the fact that vortex $k$ only moves due to the velocity field induced by all other vortices, i.e.\ the dynamics are determined entirely by $H_>$. Differentiating Eq.\ \eqref{eq:Hgreater} with respect to $\mathbf{x}_k$, and taking the vector product with $\hat{\mathbf{z}}$, finally gives the well-known equation of motion \eqref{eq:PV_EOM} for a point vortex.

\section{Discussion and conclusion}
\label{sec:discussion_conclusion}

Before we state our conclusions, we first make some remarks regarding the assumptions used in this derivation, and the order of correction in the calculation.

\subsection{Discussion}
\label{subsec:discussion}

\subsubsection{Assumption of a neutral ensemble and discussion of boundary conditions}
\label{subsubsec:neutral_ensemble}

In this work we restrict our consideration to a neutral ensemble of vortices. This is to ensure that the fluid velocity $\mathbf{u}$, and hence $\nabla\psi$, vanishes at infinity (assisted by $\rho\to 1$ in that limit), allowing us to integrate by parts with a vanishing boundary term contribution at infinity. 
	As mentioned in Sec.~\ref{sec:transformingH}, the energy calculations by other authors~\citep{pitaevskii1961vortex, ginzburg1958theory, maestriniPhDthesis}, of non-neutral vortex ensembles in finite discs of radius $R$, lead to non-vanishing boundary contributions (dependent on the specific boundary conditions imposed), which cannot necessarily be neglected in the $R\to\infty$ limit. 
	In those works, the authors implicitly assume a free-slip boundary at radius $R$. 
	In that case, the leading-order contribution to the energy comes from the monopole moment due to the imbalance of positive and negative vortices, and diverges as $\sim \! \log R$. 
	
	Here we are interested in the infinite system, and so formally need to take $R\to\infty$, before differentiating the Hamiltonian on the RHS of Eq.\  \eqref{eq:dHdxk_penultimate}, to get the PV model. However, the differentiation cannot be done when $H$ contains a divergent contribution. By contrast, for the neutral ensemble the monopole moment vanishes, and so the boundary term decays to zero as $R\to\infty$. Thus, in order to maintain a level of rigour that is commensurate with the rest of this derivation, we restrict ourselves to a neutral ensemble of vortices in the infinite system.
	
	Having said that, our derivation can be modified to a non-neutral collection of vortices inside a \emph{finite} disc, with free-slip boundary conditions, as follows. 
	By uniqueness of solutions \eqref{eq:Psi_field} to Poisson's equation, a free-slip boundary can be reproduced by the method of images. 
	Each real vortex within the disc will have a single image vortex of opposite sign, lying outside the disc. 
	The overall ensemble of real and image vortices will therefore be neutral, and the entire derivation of this paper follows, with the boundary term of the partial integration taken at infinity (i.e.\ beyond the finite radius of the disc). 
	The only caveat is that the PV equations of motion \eqref{eq:PV_EOM} apply only to the real vortices within the disc, and do not apply to the image vortices, as the streamfunction obtained outside the disc is purely auxiliary: the image vortices have unphysical dynamics.

	Finally, we remark that choosing a boundary other than the disc requires additional care, particularly in situations where reproducing the boundary conditions requires an infinite lattice of replica vortices.
	Such an infinite series of replicas pertains to systems with periodic boundary conditions \citep{campbell1989energy}.
The problem of rectangular free-slip domains can also be mapped to the periodic problem, after reflecting the system into a two-by-two cell, to ensure periodicity of the phase \citep{maestrini2019entropy, maestriniPhDthesis}. 
	Furthermore, periodic systems of point vortices in the Euler equations were considered in Ref.\ \citep{weiss1991nonergodicity}, but it was shown that the motion of vortices in the corresponding periodic NLS system differs by a constant velocity drift, arising from the requirement of a
periodic phase with respect to the boundary \citep{griffin2020magnus}.
	Consequently, the derivation of the PV model from the NLS equation in the rectangular domain requires additional considerations that go beyond the treatment in this paper.

\subsubsection{Assumptions of well-separated vortices and incompressible flow}
\label{subsubsec:wellsep_incompr}

Our assumptions also include that of well-separated vortices, and the exclusion of the compressible (acoustic) excitations. 
	If these assumptions are relaxed, the Pitaevskii profiles of the vortices can become significantly distorted. 
	In that case, we can no longer decompose the problem into that of $N$ vortices with identical density profiles, interacting from afar. 
	The errors in the quantities that we calculate in this derivation (e.g.\ the terms in the Hamiltonian) may become of the same order as, or even exceed, the quantities themselves, meaning the PV model will no longer be a good approximation of the NLS equation.
	
The generation of acoustic excitations and the close proximity of vortices go hand-in-hand: it is known that in a full NLS system, acoustic waves are excited when vortices approach each other, and indeed sound is vitally important in the process of vortex annihilation \citep{nazarenko2006wave, zuccher2012quantum}.
On annihilation, the self-energy $H_\mathrm{self}/N$ of each participating vortex will be liberated into the energy of acoustic waves. The total $H_\mathrm{self}$ of the vortex collection thus provides a lower bound on the amount of acoustic energy that can be produced by annihilating all vortices (although the contributions from $H_>$ are unbounded). 
We note that vortex production, collision, and annihilation can now be manipulated with exquisite control experimentally \citep{kwon2021sound}, and our calculation of $H_\mathrm{self}$ might be used to help quantify the acoustic energy produced in such annihilations.

\subsubsection{Order of correction}

To estimate the error in the calculation, let us assume that the system stays within the assumptions stated in Sec.\ \ref{sec:assumptions}. 
	In the discussion before Eq.\ \eqref{eq:Hgreater}, the assumption of well-separated vortices allowed us to collapse the vorticity profiles to delta functions. 
	The leading correction will come from the variation of the vorticity of one vortex (say the $j$th) over the $\sim \! \xi$ width of its nearest neighbour (say the $m$th), located at distance $\ell_{jm}$. 
	Considering the nearest pair of vortices in the ensemble, we need to consider how the vorticity $\omega_j(r)$ varies between $r=\ell$ and $r=\ell + \xi$.
	Taylor expanding the vorticity profile \eqref{eq:omegaprofile} gives a relative error in $\omega_j$ of $\mathcal{O}(\xi/\ell)$.
	Propagating this error through the calculation, and noting that this is an upper bound, we see that the PV model reproduces the dynamics of the full NLS equation to within $\mathcal{O}(\xi/\ell)$, if the conditions of  Sec.\ \ref{sec:assumptions} remain adhered to.

\subsection{Conclusion}
\label{subsec:conclusion}

In this work we have derived the equation of mutually-induced motion of a collection of point vortices \eqref{eq:PV_EOM} (with the collection including the same number of positive and negative vortices) from the 2D defocusing NLS equation in its Hamiltonian formulation \eqref{eq:HamEq}.
Our approach complements previous derivations of the PV model by other methods, e.g.\ Refs.\ \citep{fetter1966vortices, neu1990vortices, maestriniPhDthesis}.

In particular, considering the short and long-range contributions to the Hamiltonian has allowed us to calculate, for the first time to our knowledge, the self-energy per vortex, $H_\mathrm{self}/N$ \eqref{eq:Hself_result}, in such a collection.

In addition, we have found that the contribution to the kinetic energy local to each vortex core, $H_</N$, is negative. However, since the total kinetic energy is manifestly positive, c.f.\ Eq.\ \eqref{eq:H_K}, this establishes a lower bound on the mutual, and hence the point vortex, energy $H_> > |H_<|$.
 
As a final mathematical remark, we note that the rigorous derivation of the PV model we present here is simplified by the fixed profiles of a Pitaevskii vortex, Eqs.\ \eqref{eq:PitaevskiiVortex_u} and \eqref{eq:PitaevskiiVortex}. By contrast, in the Euler equations, the density and vorticity profiles have considerable functional freedom. There will be no universal value for $H_\mathrm{self}/N$ in a system of hydrodynamic vortices, and the calculation of $H_\mathrm{self}$ will depend on the particular profiles of each vortex in the system.

	The dynamical stability of a single vorticity profile should also be guaranteed, before considering the limit of an array of such vortices. In our case of the NLS, the rigidity of the Pitaevskii profile provides a regularisation that gives us a delta-distributed vorticity, which we mollify by introducing the pseudovorticity, that is stable in the well-separated assumption. This advantage highlights the attractive property of the NLS equation as a mathematical regularisation of the 2D Euler equations.

\section{Acknowledgements}

This work was supported by the European Union's Horizon 2020 research and innovation programme under the Marie Sk\l odowska-Curie grant agreement No. 823937 for the RISE project HALT.
J.L. and J.S. are supported by the Leverhulme Trust Project Grant RPG-2021-014.

\bibliographystyle{unsrtnat}
\bibliography{NLS_to_PtVortex}

\end{document}